\documentclass[reprint,prb,aps,amsmath,amssymb,floatfix,showpacs]{revtex4-1}

\usepackage[usenames,dvipsnames,svgnames,table]{xcolor}
\usepackage{graphicx}% Include figure files
\usepackage{dcolumn}% Align table columns on decimal point
\usepackage{bm}% bold math
\usepackage{amsmath}
\usepackage{subfigure}
\usepackage[normalem]{ulem}
\begin{document}

\newcommand{\bo}{\boldsymbol}
\newcommand{\boq}{\mathbf{q}}
\newcommand{\bok}{\mathbf{k}}
\newcommand{\bor}{\mathbf{r}}
\newcommand{\boG}{\mathbf{G}}
\newcommand{\boR}{\mathbf{R}}
\newcommand{\comment}[1]{\textcolor{red}{#1}}

\title{Breakdown of optical phonons' splitting in two-dimensional materials}
\author{Thibault Sohier$^{1}$}
\author{Marco Gibertini$^{1}$}
\author{Matteo Calandra$^{2}$}
\author{Francesco Mauri$^{3,4}$}
\author{Nicola Marzari$^{1}$}

\affiliation{
$^{1}$Theory and Simulation of Materials (THEOS), and National Centre for Computational Design and Discovery of Novel Materials (MARVEL), \'Ecole Polytechnique F\'ed\'erale de Lausanne, CH-1015 Lausanne, Switzerland \\
$^{2}$IMPMC, CNRS, Universit\'e P. et M. Curie, 4 Place Jussieu, 
75005 Paris, France\\
$^{3}$Departimento di Fisica, Universit\`a di Roma La Sapienza, 
Piazzale Aldo Moro 5, I-00185 Roma, Italy \\
$^{4}$Graphene Labs, Fondazione Istituto Italiano di Tecnologia
}

\date{\today}


\begin{abstract}
We investigate the long-wavelength dispersion of 
longitudinal and transverse optical phonon modes 
in polar two-dimensional materials, multilayers, and their heterostructures.
Using analytical models and density-functional perturbation theory 
in a two-dimensional framework, we show that, 
at variance with the three-dimensional case,
these modes are degenerate at the zone center but the 
macroscopic electric field associated with the longitudinal-optical modes 
gives rise to a finite slope at the zone center in their 
corresponding phonon dispersions.  
This slope increases linearly with the number of layers and it 
is determined solely by the Born effective charges of the material 
and the dielectric properties of the surrounding media. 
Screening from the environment can greatly reduce the slope splitting 
between the longitudinal and transverse optical modes and can be seen in 
the experimentally relevant case of boron nitride-graphene heterostructures. 
As the phonon momentum increases, the intrinsic screening properties of 
the two-dimensional material dictate the transition to a momentum-independent splitting similar to that of three-dimensional materials. 
These considerations are essential to understand electrical transport and optical 
coupling in two-dimensional systems.
\end{abstract}

\maketitle

%%%%%%%%%%%%%%%%%%\section{Intro}
Van der Waals heterostructures \cite{Geim2013} will assuredly play a key 
role in future electronic and optoelectronic devices \cite{Fiori2014,Jariwala2014,Wang2012a,Xia2014}.
Many of the potential candidates to build those next-generation devices 
are polar two-dimensional (2D) materials,
including transition-metal dichalcogenides (TMDs) and hexagonal boron nitride (h-BN).
A significant consequence of polarity in these materials 
is the generation of long-ranged electric fields by the polarization density
associated with their longitudinal optical (LO)
phonon modes. These fields strongly
influence the transport properties of the monolayers 
as well as their heterostructures, via the Fr\"ohlich electron-phonon 
interaction~\cite{Verdi2015,Sjakste2015,Sohier2016}.
Importantly, they lead to additional 
dipole-dipole interaction terms affecting 
the dispersion characteristics of optical phonons.
This leads to a splitting between the LO and 
transverse optical (TO) modes, 
driven by the long-ranged Coulomb interactions and electronic 
screening. Here, we show in detail how this LO-TO splitting 
is drastically affected by dimensionality.
In 3D, it is well understood how this splitting is independent of 
the norm of the phonon momentum and how it lifts the degeneracy of the LO and TO 
phonon modes in the long wavelength limit.
We show here that in 2D, the splitting depends on the norm of the phonon momentum, 
vanishes at the zone center and 
leads to a discontinuity in the derivative of the LO phonon dispersion.
First-principles computations of this phenomenon are delicate 
independently of dimensionality, due to the 
long-ranged nature of the dipole-dipole interactions leading to electric fields 
that are not compatible with periodic boundary conditions.
In 3D materials, however, the development of
analytical models based on the Born effective charges 
and the dielectric tensor has allowed for the correct numerical 
treatment of the LO-TO 
splitting\cite{Kunc1982,Giannozzi1991,Baroni,Gonze1994,Gonze1997,Kern1999}.
Similar efforts remain to be attained in 2D materials.
A few theoretical and computational 
works have identified the main characteristics of the LO-TO splitting 
in 2D h-BN. Namely, it was pointed out that the splitting vanishes
at the zone center \cite{Mele2002} while the slope of the LO dispersion is finite
\cite{Sanchez-Portal2002,Michel2009,Michel2011}. 
However, the peculiarity of screening in a 2D framework was not accounted for, 
and the proposed slopes were not quantitatively accurate. 
Furthermore, the empirical force constant models used lacked the generality 
and predictive power of first-principles simulations.

It should be pointed out that in most first-principles 
calculations of two-dimensional materials, 
the 2D system is repeated periodically while using a large interlayer distance. 
In such a setup, the splitting does not vanish at the zone center, due to
the spurious long-range interactions with the periodic copies. 
The equality of LO and TO frequencies 
at zone center can be enforced by simply omitting the 3D 
splitting \cite{Zolyomi2014}, but 
the dispersion at small momenta remains erroneous due to spurious
interactions between the periodic images.
Alternatively, the authors of Ref. \onlinecite{Wirtz2003} study
zone-border phonons so that LO phonon modes in 
neighbouring layers are out-of-phase. In this case, 
the electric fields generated by periodic images
cancel out at long wavelengths. However, this approach also removes 
the effects of the electric field generated by a single isolated layer, 
and so it is not completely satisfactory.
In this work we study in detail the LO-TO dispersions in 2D materials, 
that is their slope splitting at the zone center 
as well as the transition to a flatter dispersion for the LO mode 
at larger momenta. This is accomplished by formulating 
a detailed and quantitative analytical model
that takes into account the subtle effects of screening in 2D,  
and by performing density-functional perturbation theory 
(DFPT) calculations of the LO-TO splitting in the appropriate 
open boundary 2D framework, 
using our implementation \cite{Sohier2015a,Sohier2016} of 
Coulomb cutoff techniques \cite{Ismail-Beigi2006,Rozzi2006} in the relevant codes
(PWSCF and PHONON) of the Quantum ESPRESSO distribution \cite{Giannozzi2009,Baroni}. 
This latter approach unlocks the full potential and versatility 
of DFPT methods for 2D systems, and allows us to study 
LO-TO splitting in different monolayers, 
multilayers and heterostructures, 
providing the insight needed to interpret the results of 
various experimental setups and their effect on transport measurements. 

%%%%%%%%%%%%%%%%\section{Model}
In order to frame the discussion, we begin with introducing our 
analytical model. We are interested in the dispersions of LO and TO phonons 
near the $\mathbf{\Gamma}$ point in 2D materials or layered 3D materials, 
and we consider a phonon of in-plane momentum 
$\boq_p$. For simplicity, we use the
$|\boq_p| \to 0$ limit of the phonon displacements 
and neglect the deviation from the strictly longitudinal
and transverse nature of the phonon modes as 
momentum increases. The displacement of atom $a$ in the unit cell
is then given by $\mathbf{u}^{a}_{\rm{LO}}=\mathbf{e}^{a}_{\rm{LO}}/\sqrt{M_a }$,  
where $M_a$ is the mass of atom $a$ and $\mathbf{e}^{a}_{\rm{LO}}$ is 
the $|\boq_p| \to 0$ limit of the eigenvector of the dynamical matrix 
corresponding to the LO mode, normalized over the unit cell.
The origin of the LO-TO splitting in polar materials
is the polarization density generated by the atomic 
displacement pattern $\mathbf{u}^{a}_{\rm{LO}}$ and 
the associated electric fields.
The Fourier transform \footnote{See supporting information for the 
definition of the Fourier transform depending on the 
dimensionality of the system} of the polarization density is
\begin{align} \label{eq:pola}
\mathbf{P}(\boq_p) &= \frac{e^2}{\Omega} \sum_a \bm{\mathcal{Z}}_{a} \cdot 
\mathbf{u}^{a}_{\rm{LO}},
\end{align}
where $e$ is the elementary charge and $\Omega$ is either 
the volume of a 3D-periodic system's unit cell $V$ 
or the area of a 2D-periodic system's unit cell $A$.
The tensor of Born effective charges associated with atom $a$ in the 
unit-cell is $\bm{\mathcal{Z}}_a$.
Note that in the general case, treated in Appendix, 
it is not necessarily possible to define a pair of 
LO/TO modes belonging to the same irreducible representation at 
$\mathbf{\Gamma}$. 
For simplicity and clarity, we focus here on materials 
where this is possible, such as the commonly studied materials 
with hexagonal in-plane symmetry. Further, we assume in-plane isotropy 
with respect to long-wavelength perturbations,
strictly in-plane phonon displacements and diagonal 
Born-effective-charges tensors.
Within these assumptions, valid for all 
the materials mentioned in this work, 
the LO and TO modes would be mechanically 
similar in the long-wavelength limit.
In polar materials, however, the emergence of long-ranged dipole-dipole 
interactions differentiates them.
The divergence of the polarization, with Fourier transform $\boq_p \cdot \mathbf{P}(\boq_p)$, 
represents a polarization charge density. 
This polarization charge density is zero for the TO mode 
due to the orthogonality of the polarization and 
the direction of propagation. The LO mode, however, does induce
a polarization charge density which, in turn, generates an electric field.
The Born effective charges describe the atomic response to such a field 
and this leads to an additional restoring force 
on the atoms and an increase in energy cost for the displacement pattern of the LO mode 
with respect to the TO mode. 
The resulting relationship between the square of the frequencies of these modes
is well-known in 3D bulk materials \cite{Baroni,Giannozzi1991,Gonze1997}.
As the central analytical result of this paper, 
we generalize this relationship for both 2D and 3D layered materials as
\begin{align} \label{eq:LOTOsplit}
\omega^2_{\rm{LO}}& = \omega^2_{\rm{TO}}+W_{c}(\boq_p) \frac{e^2|\boq_p|^2}{\Omega} 
\left( \sum_{a} \frac{ \mathbf{e}_{\boq_p} \cdot \bm{\mathcal{Z}}_{a} 
\cdot   \mathbf{e}^{a}_{\rm{LO}}}{\sqrt{M_a } } 
\right)^2 ,
\end{align}
where $\mathbf{e}_{\boq_p}=\boq_p/|\boq_p|$ and $W_{c}(\boq_p)$ is a screened
Coulomb interaction discussed below.
In general, the second term on the right-hand side depends on the momentum direction 
$\boq_p$ via the Born effective charges and the screening. 
Here, we focus on in-plane isotropic materials for which all quantities 
depend only on the magnitude of $\boq_p$ for $\boq_p\to0$.
We capture the fundamental role of dimensionality in the long wavelength limit
by introducing the following simple model for the screened macroscopic Coulomb 
interaction
\begin{align} \label{eq:W}
W_{c}(\boq_p)= 
\begin{cases}
\frac{4\pi }{|\boq_p|^2  \epsilon^b_p} & \text{in layered 3D materials} \\
\frac{2\pi }{|\boq_p|\epsilon_{\rm{2D}}(|\boq_p|)} & \text{in 2D materials}
\end{cases},
\end{align}
where, in addition to the different powers at which $|\boq_p|$ 
appears in the Coulomb interaction, screening is considered differently 
in 2D and 3D. 
In the 3D layered materials discussed here, screening can be described 
by the in-plane dielectric constant of the bulk $\epsilon^b_p$. 
The splitting (second term in Eq \eqref{eq:LOTOsplit}) 
is then independent of  momentum, with an expression 
that depends notably on the effective charges and the 
dielectric constant.
In 2D, screening can be described by
$\epsilon_{\rm{2D}}(|\boq_p|) = \epsilon_{\rm{ext}}
+ r_{\rm{eff}} |\boq_p|$ \cite{Keldysh1979,Wehling2011,Cudazzo2011,
Berkelbach2013,Steinhoff2014,Sohier2016}, 
where the constant $\epsilon_{\rm{ext}}$ describes the 
dielectric properties of the environment. 
In the case of two semi-infinite dielectrics on each side of the 
2D material, with relative permittivity $\epsilon_1$ and $\epsilon_2$,  
we have $\epsilon_{\rm{ext}}=\frac{\epsilon_1+\epsilon_2}{2}$.
In the case of an isolated monolayer $\epsilon_{\rm{ext}}=1$.
The effective screening length $r_{\rm{eff}}$ describes the screening 
properties of the 2D material itself. It can be approximated
as $r_{\rm{eff}}\approx \epsilon^b_p t/2$ where $t$ is 
the thickness of the 2D material (see Ref. \onlinecite{Sohier2016} for details).  
Due to the effect of dimensionality on the screened Coulomb interaction,
the 2D splitting now depends on momentum. Namely, Eq. \eqref{eq:LOTOsplit} 
can be written in the shorthand form 
$\omega^2_{\rm{LO}}\approx \omega^2_{\rm{TO}}+
\mathcal{S}|\boq_p|/\epsilon_{\rm{2D}}(|\boq_p|)$
where $\mathcal{S}$ is a constant depending on the effective 
charges and the masses. 
For $\boq_p \ll \epsilon_{\rm{ext}} r_{\rm{eff}}^{-1}$, 
the splitting is linear in $|\boq_p|$
and screened solely by the surrounding medium. 
As momentum decreases, the electric 
field lines associated with the polarization charge density 
spread more and more in the surrounding medium, 
leading to a vanishing dipole-dipole interaction and thus a vanishing 
splitting. At $\Gamma$, although the  
splitting is zero, the slope of the LO dispersion is finite and discontinuous. 
It has a positive value $\frac{\mathcal{S}}{2\epsilon_{\rm{ext}}\omega_{\rm{TO}}}$ 
in all directions.
For $\boq_p \gg \epsilon_{\rm{ext}} r_{\rm{eff}}^{-1}$, 
one recovers the 3D case. 
Indeed, in this limit, the electric field associated with the 
polarization density is confined within the thickness of the monolayer. 
The dipoles interact within the layer
and it makes no difference whether the monolayer is isolated or
surrounded by other monolayers (as in a layered 3D material). 
The material-specific effective 
screening length $r_{\rm{eff}}$ determines the transition between the two 
regimes, and can be estimated from first-principles  \cite{Wehling2011,Cudazzo2011,Berkelbach2013,Steinhoff2014,Sohier2016}.
\begin{figure}[t]
\centering
\includegraphics{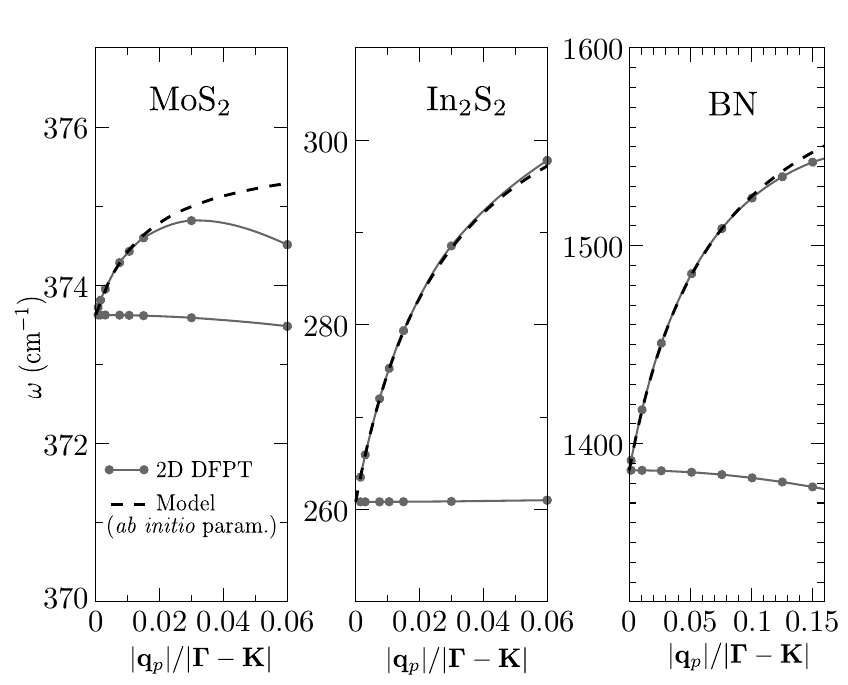}
\caption{LO-TO splitting for three different isolated monolayers. 
We show direct DFPT phonon calculations in 2D open boundary conditions 
of the LO and TO modes as well
as the model of Eq. \eqref{eq:LOTOsplit} for the LO mode. 
For the model we use the first-principles parameters of table \ref{tab:param}, and 
$\epsilon_{\rm{ext}}=1$. Phonon momenta are in the $\mathbf{\Gamma-K}$ direction.
Note that the scales are different, and the slope splitting 
strongly increases in going from MoS$_2$ to BN. }
\label{fig:model}
\end{figure} 
\begin{table}[t]
\caption{First-principles parameters related to LO-TO splitting in various polar 2D materials. 
$\omega^2_{\rm{LO}}= 
\omega^2_{\rm{TO}}+\mathcal{S}|\boq_p|/\epsilon_{\rm{2D}}(|\boq_p|)$
with $\epsilon_{\rm{2D}}(|\boq_p|) = \epsilon_{\rm{ext}}
+ r_{\rm{eff}} |\boq_p|$. The effective dielectric constant 
$\epsilon_{\rm{ext}}$
of the surrounding medium is defined as $\frac{\epsilon_1+\epsilon_2}{2}$ 
where $\epsilon_1$ and $\epsilon_2$ are the dielectric constants of the 
surrounding media on each side.
The parameter $\mathcal{S}$ is computed from first-principles 
Born effective charges and phonon displacements.
The effective screening length of the 
monolayer $r_{\rm{eff}}$ is computed via an effective medium model 
as in Ref. \onlinecite{Sohier2016}.}
\begin{tabular}{ c c c c }
\hline
Monolayer & $\mathcal{S}$ (eV$^{2}\cdot $ \AA\ ) & $r_{\rm{eff}}$ (\AA) 
& $\omega_{\rm{TO}}$ (cm$^{-1}$)  \\ 
\hline
h-BN       & $8.40\ 10^{-2}$ & $7.64$ & $1387.2$  \\
MoS$_2$  & $1.13\ 10^{-3}$ & $46.5$ & $373.7$  \\
MoSe$_2$ & $2.09\ 10^{-3}$ & $53.2$ & $277.5$  \\
MoTe$_2$ & $4.87\ 10^{-3}$ & $69.5$ & $223.6$  \\
WS$_2$   & $2.10\ 10^{-4}$ & $42.0$ & $345.9$  \\
WSe$_2$  & $6.25\ 10^{-4}$ & $48.7$ & $239.4$  \\
In$_2$S$_2$ & $1.37\ 10^{-2}$  & $28.62$ & $260.8$   \\
In$_2$Se$_2$ & $6.58\ 10^{-3}$  & $35.77$ & $179.0$ \\
\hline
\end{tabular}
\label{tab:param}
\end{table}

\begin{figure}[t]
\centering
\includegraphics{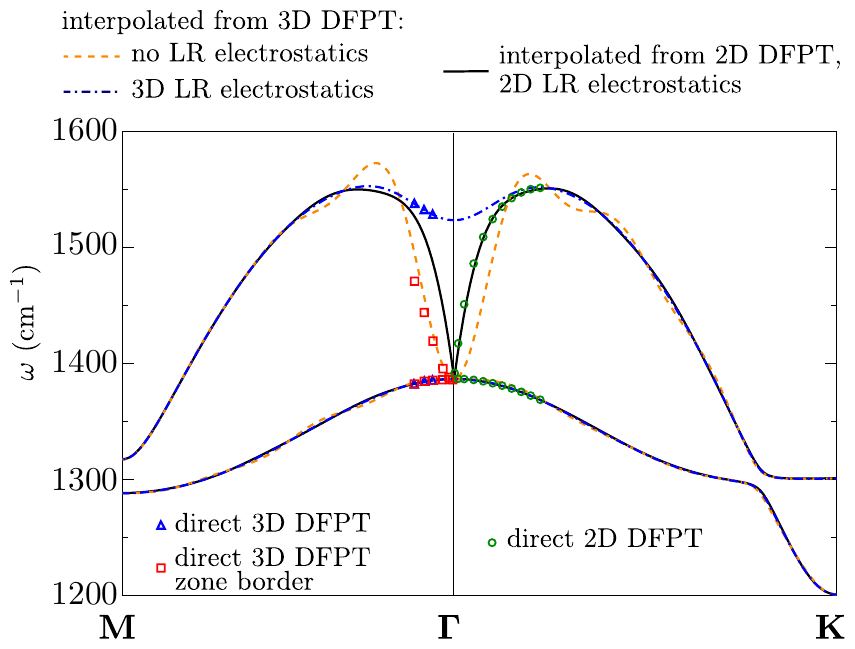}
\caption{LO-TO phonon dispersions in monolayer h-BN using various methods. 
Phonons are computed on a $12 \times 12$ in-plane grid using DFPT in 2D (with 2D open boundary conditions) 
or 3D (with $\approx 15$ \AA\ of vacuum between periodic images).
The dispersions are then Fourier interpolated, using different treatments 
for the long-range (LR) electrostatic dipole-dipole interactions. 
``No LR electrostatics'' means that we directly interpolate the result of DFPT (with vanishing splitting at $\mathbf{\Gamma}$). The 3D treament of long-ranged electrosctatics (``3D LR electrostatics'') is described in \cite{Gonze1997,Baroni,Giannozzi1991}. 
The 2D treatment (``2D LR electrostatics'') is described in the text.
For reference, we include direct single q-point calculations: green dots for 2D DFPT, 
blue triangles for 3D DFPT (with $\approx 15$ \AA\ of vacuum), and red squares for 3D DFPT using zone border phonons and $\approx 7.3$ \AA\ of vacuum as in Ref. \onlinecite{Wirtz2003}.}
\label{fig:interp}
\end{figure}

%%%%%%%%%%%%%%%%%%%%%%%% COMPUTATION
To complement these analytical results we performed DFPT calculations in 2D open
boundary conditions of the 
of LO and TO dispersions in a selection of monolayers, as shown
in Fig. \ref{fig:model}. 
These are in excellent agreement with the 2D analytical model of 
Eq. \eqref{eq:LOTOsplit}, using the parameters obtained independently 
from  first principles and reported in Table \ref{tab:param}.
Note that at larger momenta the LO dispersion displays some
material-specific behaviour, as seen more clearly for MoS$_2$.
Contrary to the 3D case, the computation of 
the frequency of the LO mode is not problematic at $\Gamma$ exactly, 
since the splitting is zero.
To obtain the correct behaviour at small but finite momentum, 
however, several issues arise.
In standard plane-wave DFPT, spurious interactions with the artificial 
periodic images yield erroneous results in the long-wavelength limit: 
when the phonon momentum is smaller than the inverse of the distance 
between periodic images, 
the atoms of one monolayer feel the polarization field of the 
neighbouring monolayers. In addition, the periodic images 
lead to spurious screening\cite{Kozinsky2006,Sohier2015}. 
No matter the amount of vacuum inserted in between the periodic images, 
for sufficiently small momenta one will always end up with the response
of a 3D-periodic system, i.e. a non-vanishing splitting, as shown in Fig.~\ref{fig:interp}. 
This issue is solved by using our implementation of the 2D Coulomb cutoff
technique \cite{Sohier2015a,Sohier2016}, as shown in Fig. \ref{fig:model}.
In addition, a technical difficulty arises when using Fourier 
interpolation schemes \cite{Gonze1997,Baroni,Giannozzi1991}, 
that are otherwise very useful to obtain phonon dispersions on fine momentum 
grids at minimal computation cost.
In fact, the discontinuity in the first derivative of the LO dispersion, 
due to the long-ranged dipole-dipole interactions,
leads to interatomic force constants 
(IFC) decaying with power-law in real-space \cite{Piscanec2004}. 
Since the Fourier interpolation scheme relies on finite-ranged 
IFCs, it is not suited to capture correctly LO-TO splitting. 
A similar issue is present in 3D, with a discontinuity in 
the value of the frequency rather than in its derivative. 
In 3D, the established
solution \cite{Kunc1982,Giannozzi1991,Baroni,Gonze1994,Gonze1997,Kern1999} 
is to construct a model of the long-ranged dipole-dipole interactions 
in reciprocal space, such that the corresponding contribution to the 
dynamical matrices can be excluded from the interpolation process.
We apply the approach detailed in Ref. \onlinecite{Gonze1997} 
to the 2D case, simply replacing the 3D screened Coulomb interaction with the 2D one (for the case of isotropic materials it is given in Eq. \eqref{eq:W}, 
for the anisotropic materials it is in the Appendix).
In the long-wavelength limit, this amounts to excluding from the interpolation 
the following contribution to the dynamical matrix :
\begin{align}
\mathcal{D}^{\bm{\mathcal{Z}}}_{ai,a'j} (\boq_p\to0)&=  \frac{e^2}{\Omega} \frac{2\pi}{|\boq_p|}
\frac{\left( \boq_p\cdot \bm{\mathcal{Z}}_{a} \right)_i 
 \left(\boq_p \cdot \bm{\mathcal{Z}}_{a'}\right)_j 
}{\sqrt{M_a M_{a'}}} 
\end{align}
where $i,j$ are Cartesian coordinates.  
In principle, the long-wavelength 
\textit{bare} dipole-dipole interaction above is sufficient to treat 
the non-analyticity of the dynamical 
matrix giving rise to the power-law decay of IFCs. 
In practice, we also include screening at finite momenta via the use of the 
screened 2D Coulomb interaction. This helps convergence with 
respect to the phonon momentum grid from which the dispersions are interpolated.
Indeed, screening brings some analytical but potentially sharp variations 
to the LO dispersion.
Fig. \ref{fig:interp} shows that this interpolation scheme is successful 
in reproducing the correct long-wavelength behavior.
To highlight the importance of adapting DFT and DFPT to 2D materials,
we also show how standard 3D periodic boundary conditions calculations fail in describing the long-wavelength behavior of 
polar-optical phonons in 2D.

%%%%%%%%%%%%%%%%%% MULTILAYERS

\begin{figure}[t]
\includegraphics{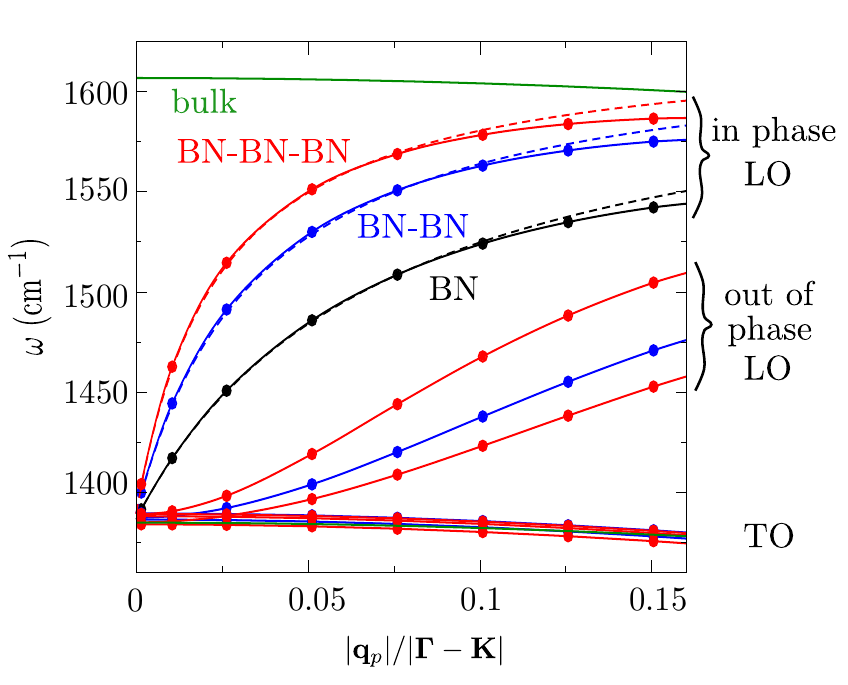}
\caption{Evolution of LO-TO splitting in multilayer BN as a function of
the number of layers. 
We show direct 2D DFPT phonon calculations 
(plain lines with dots corresponding to data points)
as well as the results of the model of Eq. \eqref{eq:multi} (dashed lines).
Phonon momenta are in the $\mathbf{\Gamma-K}$ direction.
The layer separation is set to the experimental value for bulk BN 
($\approx 3.25$ \AA). For reference, we also show the bulk limit, in green, 
computed with standard 3D DFPT.}
\label{fig:multi}
\end{figure}
For a comprehensive understanding of 2D systems 
we also study and propose a simple model of LO-TO splitting in multilayers. 
We take here h-BN as an example. 
Van-der-Waals interactions are neglected. 
These might affect the absolute value of the LO and TO frequencies by 
influencing interlayer distances but they would not change the LO-TO 
splitting itself since it is an electrostatic effect depending on in-plane 
Born effective charges and screening.
We assume that, aside from thicker slabs, 
the dielectric properties of the multilayers remain unchanged.
The underlying assumption is that the perturbing field and the material's 
response to it are uniform over the multilayer.
Furthermore, we focus on the the highest LO 
branch, with an in-phase LO mode in each monolayer, noted here as ``hLO''.
We find the frequency of this mode to be 
\begin{align}\label{eq:multi}
\omega^2_{\rm{hLO}}=  \omega^2_{\rm{hTO}}+ 
\frac{N \mathcal{S} |\boq_p|}{(1+Nr^{mono}_{\rm{eff}}|\boq_p|)}.
\end{align}
where the index ``hTO'' designates the highest TO mode 
(see Appendix for details).
Both the strength of the splitting and the screening length 
are multiplied by a factor $N$ with respect to the monolayer. 
We compare the model of Eq. \eqref{eq:multi} 
with our DFPT calculations in Fig. 
\ref{fig:multi}, showing excellent agreement for the slope at $\Gamma$ and its evolution. 
Moving away from $\Gamma$, the splitting is  
dictated by the screening of the 2D material and the model is in very good 
agreement considering the assumptions made
about the dielectric properties of the multilayer. 
The model of Eq. \eqref{eq:multi} only treats the highest LO mode of the 
multilayer, for which the polarization densities are in-phase and 
the splitting effect from each layer adds up 
constructively. 
The lower LO modes, noted ``out-of-phase LO'' in Fig. \ref{fig:multi}, 
behave quite differently. In particular, 
they do not display a finite slope at zone center.
Indeed, the electric fields generated by the polarization densities 
in different layers cancel each other 
in the long-wavelength limit, due to the atomic displacements
being out-of-phase. The same argument explains the absence of splitting
at zone center for the lower TO mode in bulk h-BN with AB stacking.

The expression in Eq.~\eqref{eq:multi} offers further insights 
on the 2D/3D transition.
As the number of layers $N$ increases, the slope of the LO mode increases 
and the range of momenta over which the dispersion is linear decreases.
For increasing $N$ the slope of the highest LO mode becomes very large, 
but the range of momenta for which the frequency increases closes up around 
$\mathbf{\Gamma}$ \footnote{Note that if one could probe phonons closer and closer 
to $\mathbf{\Gamma}$ as $N\to \infty$, the slope (or group velocity) would approach 
and surpass the speed of light, and the non-relativistic electrostatic framework 
used here would break down.}
. Eventually, we arrive to a situation where all physically
relevant momenta are such that $\boq > (Nr^{mono}_{\rm{eff}})^{-1}$.
In this bulk limit ($N\to \infty$), the difference in the squared frequencies 
approaches a constant value, giving rise to a finite LO-TO splitting close to 
$\mathbf{\Gamma}$,
which can be estimated to be $\frac{\mathcal{S} }{r_{\rm{eff}}}$ by considering 
$N\to \infty$ in Eq.~\eqref{eq:multi}\footnote{More rigorously, one should rather 
consider the limit
$N \to \infty$ and $|\boq_p| \to 0$, keeping the product of the two constant}.
In BN, and using Table \ref{tab:param}, we obtain 
$\frac{\mathcal{S}}{r_{\rm{eff}}} \approx 0.011$ eV$^2$. This value is within
$10 \%$ of the bulk splitting obtain in 3D DFPT (see Fig. \ref{fig:multi}), 
and in excellent agreement with experimental results \cite{Rokuta1997,Geick1966} 
in bulk h-BN. Such agreement points to a relatively easy way to estimate 
this quantity from experiments or bulk calculations.

%%%%%%%%%%%%%%%%%%SUBSTRATE
\begin{figure}[h]
\includegraphics{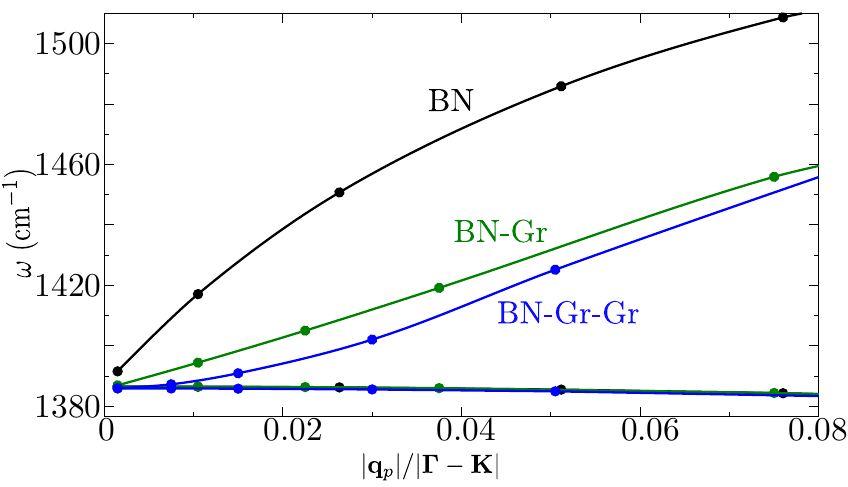}
\caption{Long-wavelength behaviour of the LO-TO splitting for isolated h-BN, monolayer h-BN on monolayer graphene and monolayer h-BN on bilayer graphene. 
Phonon momenta are in the $\mathbf{\Gamma-K}$ direction. The slope of h-BN's 
LO mode depends on the dielectric environment.}
\label{fig:BNGr}
\end{figure}
Last, in order to assess the role of environmental screening 
on LO-TO phonons we consider van-der-Waals heterostructures.
These considerations are also relevant for transport properties 
of heterostructures, as 
the LO dispersion slope is representative of the strength 
of the long-ranged polarization fields that 
can scatter electrons remotely in all layers.
In practice, it is essential to consider that 
the slope of the LO dispersion at the zone center is divided by 
the effective dielectric constant of the environment 
$\epsilon_{\rm{ext}}$.
Isolated monolayers are not always easily 
fabricated \cite{Rokuta1997,Serrano2007}, 
and future devices will rely on Van-der-Waals heterostructures, 
in which the polar 2D material is surrounded by a variety of other 2D layers.
We study the effect of the dielectric environment 
by simulating h-BN on top of monolayer and bilayer graphene (see Fig. 
\ref{fig:BNGr}). 
The calculations are performed with an electronic smearing equivalent to room 
temperature. This smearing is small enough to consider monolayer graphene as 
neutral, with a constant static screening function \cite{Kozinsky2006,Shung1986,Gorbar2002,Ando2006,Wunsch2006,Barlas2007,Wang2007a,Hwang2008}, namely $\epsilon_{\rm{gr}}\approx 5.5$ \cite{Kozinsky2006,Sohier2015}.
At small momenta, the graphene layer then behaves like an insulating bulk 
dielectric and the slope of the LO dispersion is divided by $\epsilon_{\rm{ext}}=\frac{1+\epsilon_{\rm{gr}}}{2}$.
Bilayer graphene has a larger density of states 
at the intersection of the valence and conduction bands. 
In this case the smearing is enough for the 
bilayer graphene to exhibit a metallic behaviour 
in the long wavelength limit. 
The slope of the LO dispersion vanishes, as the polarization field 
from h-BN is completely screened by the electrons of bilayer graphene.
Finally, note that in the general case of heterostructures containing other polar materials, 
the LO-TO splitting will be the result of a complex interplay between the polarization densities and screening properties of the various layers.

%%%%%%%%%%%%%%%%%%%%%%%%% CONCLUSION
In conclusion, we use our implementation of the 2D Coulomb cutoff 
within density-functional perturbation theory to study 
the long wavelength limit of polar-optical phonons in a
two-dimensional framework and complement this with physical, analytical
models of the interactions and their screening. 
At the zone center, although the splitting of LO and TO phonon modes vanishes, 
a discontinuity appears in the slope of the LO phonon dispersion, and 
we provide a model to evaluate this slope in various situations.
For isolated 2D materials, the slope can be
estimated directly from the Born effective charges.
For a multilayer, we find that the slope of the highest LO mode
is proportional to the number of layers.
In the general case, the slope also depends on 
the dielectric environment of the 2D material.
In the experimentally relevant case of h-BN/graphene heterostructures, 
the slope is reduced, and can even vanish when  
screened by the metallic behaviour of the electrons.
Last, screening from the electrons of the 2D material occurs only
for phonon wavelengths smaller than an effective 
screening length. A wavelength-independent splitting similar
to bulk 3D materials is then recovered.

%%%%%%%%%%%%%%%Acknoedgements
{\it Acknowledgements } --
This project has received funding from the European Union's Horizon 2020 research and innovation programme under grant agreement No. 696656 – GrapheneCore1.
T.S., M.G., and N.M. acknowledge financial support from the 
Swiss National Science Foundation (SNSF -- project number 200021-143636). 
M.C. acknowledges support from the Agence Nationale de la Recherche under 
the reference no ANR-13-IS10-0003-01. Computer facilities were provided
by the PRACE project and CINES, IDRIS and CEA TGCC (Grant EDARI
No. 2016091202).

\appendix
\section{Computational details}

Phonon calculations are performed within density functional perturbation
theory using a modified version of the PWSCF and PHONON codes of the Quantum ESPRESSO distribution. In particular, the modification includes the implementation 
of the 2D Coulomb cutoff for the calculation of total energy, forces, bands 
and the linear response of the system to a phonon perturbation. 
We use pseudopotentials from the Standard Solid-State Pseudopotentials (SSSP) library \footnote{see http://www.materialscloud.org/sssp} (accuracy version), 
with the exception of BN, for which we use 
ultrasoft pseudopotentials within the local density approximation.
We use $16 \times 16 \times 1$ electron-momentum grids, except when graphene is involved in which case the grid is $64 \times 64 \times 1$.
Calculations were performed using the AiiDA materials informatics platform \cite{Pizzi2016}.

\section{LO-TO splitting in the non-isotropic case}
The expression for the polarization density induced by a given phonon mode $\nu$
in a single layer is: 
\begin{align} \label{eq:pola}
\mathbf{P}(\bor_p,z) &=  \frac{e^2}{A} \sum_{a} 
\bm{\mathcal{Z}}_{a} \cdot 
\mathbf{u}^{a}_{\nu} f(z) \ e^{i \boq_p \cdot \bor_p} 
\end{align}
where $\bor_p$, $z$ are respectively in plane and out-of-plane space variables, and $f(z)$ is the out-of-plane profile of the monolayer, homogeneous to an inverse distance and normalized to unity 
($\int_{-\infty}^{\infty} f(z)dz=1$). Like in the main text, we use the 
$|\boq_p| \to 0$ limit of the atomic displacement pattern $\mathbf{u}^{a}_{\nu}$. 
The polarization density $\mathbf{P}(\bor_p,z)$ induces a potential
$V(\bor_p, z)=
V(\boq_p, z)e^{i\boq_p \cdot \bor_p}$ 
with the same periodicity, given by Poisson equation
\begin{align} \label{eq:poisson}
\nabla \cdot (\bm{\epsilon}(z) \cdot \nabla V(\bor)) 
&=4 \pi  \nabla \cdot \mathbf{P}(\bor),
\end{align}
where $\bm{\epsilon}(z)$ is a dielectric tensor depending on $z$.
Although we preserve generality in the notation of phonon mode $\nu$, 
note that a field is effectively generated only if the divergence of the 
polarization is non-zero (which is true for optical modes with a 
longitudinal component). 
To solve Eq. \ref{eq:poisson}, it is convenient to work in reciprocal space. In the context of 2D materials, since there is a periodicity only in the plane, we work with in-plane Fourier Transforms.
However, the third dimension must be treated with care to account for screening properly. In the following,
the subtleties of dimensionality and the treatment of the out-of-plane direction will be hidden in the screening. 
The in-plane Fourier Transform of $\mathbf{P}(\bor)$ is defined as follows.
In a 2D-periodic framework, we integrate over the out-of-plane variable
$\mathbf{P}(\boq_p)=\int \mathbf{P}(\boq_p,z)dz$.
The potential induced by a slab of polarization density extends much further
than its source in the out-of-plane direction, as it decays as $e^{-|\boq_p||z|}$. This calls for a slightly different definition of the in-plane Fourier Transform in the 2D framework. 
Since we are only interested in the value of the potential 
within the material, we define the in-plane Fourier Transform 
of the potential as $V(\boq_p)=\int V(\boq_p,z)f(z)dz$ in 2D. 
Note that $f(z)$ is normalized to unity.
In the 3D-periodic framework, we work with the usual average over the unit cell  $g(\boq_p)= \frac{1}{c}\int_{-c/2}^{c/2} \mathbf{P}(\boq_p,z) dz$ and
$V(\boq_p)=\frac{1}{c} \int V(\boq_p,z) dz$, where $c$ is the size of the unit cell in the out-of-plane direction.

Given those definitions, the Poisson equation is solved by:
\begin{align} 
V(\boq_p)&= W_{c}(\boq_p) \boq_p \cdot \mathbf{P}(\boq_p)
\end{align}
where $W_{c}(\boq_p)$ is the screened Coulomb interaction.
Assuming in-plane isotropy for the dielectric properties of the material, we can write it as in the main text:
\begin{align}
W_{c}(\boq_p)= 
\begin{cases}
\frac{4\pi }{|\boq_p|^2  \epsilon^b_p} & \text{in 3D} \\
\frac{2\pi }{|\boq_p|\epsilon_{\rm{2D}}(|\boq_p|)} & \text{in 2D}
\end{cases}	
\end{align}
with $\epsilon_{\rm{eff}}(|\boq_p|) = \epsilon_{\rm{ext}}
+ r_{\rm{eff}} |\boq_p|$, and $\epsilon^b_p$ is the in-plane dielectric constant 
of the bulk. See the appendix of Ref. \onlinecite{Sohier2016} for more details.
In the anisotropic case, $\epsilon^b_p$ and $r_{\rm eff}$ become tensors and we use the following model in the long-wavelength limit:
\begin{align}\label{eq:Wnoniso}
W_{c}(\boq_p)= 
\begin{cases}
\frac{4\pi }{ \boq_p \cdot  \tensor{\epsilon}_b \cdot \boq_p  } & \text{in 3D} \\
\frac{2\pi }{|\boq_p| \left( \epsilon_{\rm{ext}}
+ \frac{\boq_p \cdot \tensor{r}_{\rm{eff}}\cdot \boq_p}{|\boq_p|^2} |\boq_p|\right)}
& \text{in 2D}
\end{cases} 
\end{align}
Each component of $\tensor{r}_{\rm eff}$ can be obtained separately with the same method as in the isotropic case~\cite{Sohier2016}.

The associated electric field is:
\begin{align} 
\bm{\mathcal{E}}(\boq_p)=-\nabla V(\boq_p) 
&=- W_{c}(\boq_p) \left(\boq_p \cdot \mathbf{P}(\boq_p)\right) \ \boq_p
\end{align}
The corresponding force on atoms $a$ is:
\begin{align} 
\mathbf{F}_a &= \bm{\mathcal{E}} \cdot \bm{\mathcal{Z}}_{a}  \\
\mathbf{F}_a &= - W_{c}(\boq_p) 
\left(\boq_p \cdot \frac{e^2}{\Omega} \sum_{a'} 
\bm{\mathcal{Z}}_{a'} \cdot \mathbf{u}^{a'}_{\nu} \right)
\left( \boq_p\cdot \bm{\mathcal{Z}}_{a} \right) 
\end{align}
This bring the following additional term to the dynamical matrix:  
\begin{align}
\mathcal{D}^{\bm{\mathcal{Z}}}_{ai,a'j}&=-\frac{1}{\sqrt{M_a M_{a'}}}\frac{\partial F_{a,i}}{\partial u_{a'j}} \\
\mathcal{D}^{\bm{\mathcal{Z}}}_{ai,a'j} &=  \frac{e^2}{\Omega} W_{c}(\boq_p)
\frac{\left( \boq_p\cdot \bm{\mathcal{Z}}_{a} \right)_i 
 \left(\boq_p \cdot \bm{\mathcal{Z}}_{a'}\right)_j 
}{\sqrt{M_a M_{a'}}} 
\end{align}
Where the index $\bm{\mathcal{Z}}$ indicates that this is only the contribution related to Born-effective charges.
After selecting the eigenvalue $\omega^2_{\nu}$ by multiplying left and right 
by (the $|\boq_p| \to 0$ limit of) a generic eigenvector $\mathbf{e}_{ \nu}$, we obtain that the 
frequency is increased by:
\begin{align}
\Delta \omega^2_{\nu}(\boq_p)& = \sum_{a,i,a',j}
\frac{e^2}{\Omega} W_{c}(\boq_p)
\frac{ \left( \boq_p\cdot \bm{\mathcal{Z}}_{a} \right)_i e^{a,i}_{\nu}
 \left(\boq_p \cdot \bm{\mathcal{Z}}_{a'}\right)_j e^{a',j}_{\nu}}
{\sqrt{M_a M_{a'}}}\\
\Delta \omega^2_{\nu}(\boq_p)& = \sum_{a,a'}
\frac{e^2}{\Omega} W_{c}(\boq_p)
\frac{  \left(\boq_p\cdot \bm{\mathcal{Z}}_{a} \cdot \mathbf{e}^{a}_{ \nu} \right)
\left( \boq_p \cdot \bm{\mathcal{Z}}_{a'} \cdot \mathbf{e}^{a'}_{\nu}\right)}
{\sqrt{M_a M_{a'}}} \\
\Delta \omega^2_{\nu}(\boq_p)& = \frac{e^2}{\Omega} W_{c}(\boq_p) 
\left( \sum_{a}
\frac{  \boq_p\cdot \bm{\mathcal{Z}}_{a} \cdot \mathbf{e}^{a}_{\nu}}
{\sqrt{M_a}} \right)^2
\label{eq:splitting}
\end{align}
with $\Omega=A$ or $V$ depending on the dimensionality.  
Here we have made no assumption on the dependency of 
phonon eigenvectors or Born effective charges on 
the direction of the phonon momentum. 
Using the LO phonon eigenvector in a material with hexagonal 
in-plane symmetry like in the main text, we have 
$\Delta \omega^2_{\rm{LO}}(\boq_p)=\omega^2_{\rm{LO}}(\boq_p)-\omega^2_{\rm{TO}}(\boq_p)$.
In more general cases of 2D materials with lower symmetry, 
it is not always possible to define a pair of LO/TO modes that belong to the 
same irreducible representation at $\bm{\Gamma}$. In this case we would have
$\Delta \omega^2_{\nu}(\boq_p)=\omega^2_{\nu}(\boq_p)-\omega^2_{\nu}(\bm{\Gamma})$

\section{Evolution with the number of layers}
To compute the splitting of the highest LO branch (hLO) in multilayers, 
we first assume that the effective charges $\bm{\mathcal{Z}}_a$ are unchanged.
We then notice that, with respect to the single layer, the squared term on 
the right-hand side of Eq. \ref{eq:splitting} is multiplied by a factor 
$\left(\frac{N}{\sqrt{N}}\right)^2=N$, where the numerator comes from the 
sum over the atoms while the denominator comes from the normalization of 
the phonon eigenvectors.

For the screening, we have the same Coulomb interaction, but with
$\epsilon^{multi}_{\rm{eff}}(|\boq_p|) = \epsilon^0_{\rm{eff}}
+ r^{multi}_{\rm{eff}} |\boq_p|$ and 
$r^{multi}_{\rm{eff}}=N r^{mono}_{\rm{eff}}$ because $r_{\rm{eff}}$
is proportional to the thickness of the 2D material\cite{Sohier2016}.
We thus have:
\begin{align}
\omega^2_{\rm{hLO}} - \omega^2_{\rm{hTO}}= & 
\frac{2\pi \times N}{(1+Nr^{mono}_{\rm{eff}}|\boq_p|)} 
\frac{e^2|\boq_p|}{A} \\ 
&  
\left( \sum_{a} \frac{ \mathbf{e}_{\boq_p} \cdot \bm{\mathcal{Z}}_{a} 
\cdot   \mathbf{e}^{a}_{\boq_p \rm{LO}}}{\sqrt{M_a } } 
\right)^2 \nonumber
\end{align}
Where the squared term on the right-hand side is summed only on the atoms 
of one layer. Thus, it has the same value as in the monolayer case and all 
the consequences of having several layers is explicitly conveyed by the
presence of the $N$ factors.

\bibliographystyle{apsrev4-1}
\bibliography{LOTO}

\end{document}